\documentclass[final,3p,times]{elsarticle}

\usepackage{amssymb}

\journal{}

\begin{document}

\begin{frontmatter}



\title{Future urban mobility as a bio-inspired collaborative system of multi-functional autonomous vehicles}


\author[inst1]{Naroa Coretti Sanchez}
\author[inst2]{Juan Mugica Gonzalez}
\author[inst1]{Luis Alonso Pastor}
\author[inst1]{Kent Larson}

\affiliation[inst1]{organization={MIT Media Lab},
             addressline={},
             city={Cambridge},
             postcode={},
             state={MA},
             country={United States}}



\begin{abstract}
The fast urbanization and climate change challenges require solutions that enable the efficient movement of people and goods in cities. We envision future cities to be composed of high-performing walkable districts where transportation needs could be served by fleets of ultra lightweight shared and autonomous vehicles. A future in which most vehicles would be autonomous creates a new paradigm for the possible interactions between vehicles. Natural swarms are a great example of how rich interactions can be; they can divide tasks, cluster, build together, or transport cooperatively. The field of swarm robotics has translated some of the behaviors from natural swarms to artificial systems, proving to make systems more flexible, scalable, and robust. Inspired by nature and supported by swarm robotics, this paper proposes a future mobility in which shared, electric, and autonomous vehicles would be multi-functional and behave as a collaborative system . In this future, fleets of multi-functional vehicles would complete different tasks collaboratively, giving a response to the different urban mobility needs. This paper contributes with the proposal of a framework for future urban mobility that integrates current research and mobility trends in a novel and unique way.

\end{abstract}




\end{frontmatter}







\section{Introduction} \label{introduction}

The transportation sector has been the largest contributor to greenhouse gas emissions in the US since 2017 \cite{EPA2021}; it accounted for a 29 \% of the emissions in 2019 and, despite the measures taken to reduce the environmental impact of mobility, transportation-related emissions are still increasing \cite{EPA2021}. Moreover, predicted global urbanization and population growth rates \cite{UNreport2018} further confirm the urgency to find mobility solutions that provide an efficient and ecological service in our cities. 
 
The main drivers for change in the urban mobility landscape in the coming years are predicted to come from \textit{electrification, vehicle sharing, and autonomy} \cite{McKinsey2019}.  From the climate change perspective, vehicle \textit{electrification} combined with a deep decarbonization of the grid is predicted to be key to meet the mid-century emission reduction targets \cite{alarfaj2020decarbonizing}. In fact, many countries- United States, Japan, India, China, and most countries in the European Union - have set ambitious electrification goals for 2050 \cite{zhang2020role}. \textit{Vehicle-sharing} is expected to reduce the number of vehicles on the roads \cite{martinez2015urban,martin2010impact,millard2005car} by offering a service that combines the convenience of private vehicles without the costs and responsibilities of ownership \cite{shaheen1999short}. Vehicle sharing can also promote the use of electric vehicles, which have a high purchasing cost but low operating costs \cite{he2017service,jones2019contributions}. Finally, \textit{autonomy} is expected to increase efficiency of fleets and,  while it could also increase travel demand, research indicates that if combined with vehicle sharing, autonomy would reduce overall emissions through increased efficiency and vehicle utilization rates \cite{greenblatt2015autonomous,narayanan2020shared}.

These trends are expected to alleviate many of the urban mobility-related problems \cite{cervero2017mobility,greenblatt2015autonomous,jones2019contributions,sanchez2020autonomous}. Nevertheless, studies suggest that this might not be enough in order to meet climate change mitigation target and that a total reduction in vehicle miles traveled will be critical \cite{alarfaj2020decarbonizing, milovanoff2020electrification}. For instance, according to Alarfaj et al.\cite{alarfaj2020decarbonizing}, in order to meet a 80 \% carbon emission reduction target by 2050 with the expected travel demand, the electricity grid would have to be zero-carbon by 2050. However, if the vehicle-miles traveled were reduced to a third that carbon targets could be met an electricity carbon intensity of 100 g CO2/kWh, which is a quarter of the current carbon intensity in the US, but still above the emissions from other countries such as Sweden, France, Finland, or Austria \cite{alarfaj2020decarbonizing,EIACarbonIntensity, EEA}.

Vehicle miles traveled depend on human behavior, and this behavior depends on the built environment. Studies have shown that destination accessibility and job-house balance can reduce the vehicle miles traveled by car \cite{ewing2010travel}. This is why at our research group we advocate for future cities to have a human scale, being composed of dense and diverse high-performing districts where all the resources and amenities needed in daily life can be found within a walkable distance, drastically reducing the need for commuting \cite{alonso2018cityscope}.  In these districts, most of the trips would be one-person, short distance, and low speed, reducing car dependency and favoring more sustainable modes such as walking and shared micro-mobility \cite{alonso2018cityscope, grignard2018cityscope, lin2021affordable}. Other mobility needs such as package and goods delivery would also be served at a local neighborhood scale by fleets of shared, electric, and autonomous micro-mobility systems \cite{yao2019idk}. 

In this scenario autonomy could open many new opportunities, especially regarding the interactions between vehicles and how these interactions will impact system-level behavior. Autonomy will provide vehicles with the ability to communicate with other vehicles, humans, and the infrastructure, as well as the intelligence to make decisions based on this communication \cite{harding2014vehicle}. In this paper we argue that these communications skills open the door to a future in which mobility would work as a collaborative system. In this sense, vehicle platooning can be seen as the first step towards joint mobility, since it studies the coordinated movement of autonomous vehicles to improve safety and reduce congestion and fuel consumption \cite{maiti2017conceptualization}. However, we imagine a future in which the interaction between vehicles is much richer than communicating to coordinate.  In nature there are many inspiring examples of collaboration; for instance, in insect colonies local and simple interactions lead to complex system-level behaviors \cite{bonabeau1999swarm}. The behavior of these natural swarms has been transferred to artificial systems by researchers working in the field of swarm robotics, proving to make systems more flexible, scalable, and robust \cite{Sahin2004SwarmRF}. 

Therefore, inspired by nature and supported by swarm robotics, this paper proposes a future mobility in which shared, electric, and autonomous vehicles would be \textit{multi-functional} and behave as a \textit{collaborative system} in high-performing walkable districts. Fleets of multi-functional vehicles would complete different tasks collaboratively and giving a response to the different mobility needs of the city. The local and direct interactions between autonomous vehicles in this article has been named \textit{vehicle clustering}. When vehicles \textit{cluster}, they connect either physically or virtually, being able to share data, provide services, share computational power, transfer energy \cite{gonzalez2021dynamic} or even potentially transfer cargo.

This paper does not intend to provide all the answers to how future urban mobility will operate, but rather contribute with a framework for future mobility based on current research and mobility trends. While the presented proposal borrows from current research, to the best of our knowledge, it is the first time that such concepts are integrated this way.  

The remainder of the paper is organized as follows. Section \ref{key-ingredients} provides a description of the three key ingredients of the future that is proposed in this paper: system behavior, multi-functionality and collaboration. Section \ref{from-swarms-and-platooning} provides an overview of the fields of swarm-robotics and vehicle platooning, explaining how these concepts might be transferred to collaborative mobility as well as giving an outline of the potential benefits. Section \ref{towards-this-future} aims to convey our interpretation of the future mobility framework presented in this paper and how these concepts can be applied to the main urban mobility needs: the movement of people, movement of goods, and utility services. Finally, the main conclusions of this work are highlighted in Section \ref{conclusions}.

\section{Key ingredients for a future collaborative urban mobility}\label{key-ingredients}

We envision future cities to be composed of high performing walkable districts, in which mobility will be provided by fleets of \textit{multi-functional} shared, electric, and autonomous vehicles that behave as a \textit{collaborative system}. In this future, vehicles are regarded as parts of a system rather than as individual entities. Within the system, the relationships and interactions between vehicles are based on collaboration.

This section describes the three key ingredients for this future: system behavior, multi-functionality, and collaboration.

\subsection{System behavior}\label{system-behavior}

The cardiovascular system is an excellent semantic figurative metaphor for urban mobility \cite{cruz2015wrongfully}, and metaphors can provide us of new ways of thinking about old problems. As described by Stefanovska and Bracic:

 \begin{quotation}
In the course of evolution, individual cells organized into cellular systems of increasing complexity... At this level of organization, cells were no longer capable of individually sustaining autonomous life. A collective system that provides and distributes oxygen and nutrient materials to each cell and takes away the products of their metabolism became essential... \cite[p.~31]{stefanovska1999physics}.
 \end{quotation}

Similarly, as our cities become increasingly complex, individuals are no longer autonomous, and there is a need for a collective system, that in the case of cities, is urban mobility. This parallelism inspires us to think of urban mobility as a collective system instead of a conglomerate of people that follow their own preferences and priorities. 

A successful design and planning of urban mobility require understanding all of its layers and their interactions. Therefore, we believe that mobility - and future mobility - should not be regarded from an individualistic point of view but rather as a system. All the vehicles that are part of the urban mobility landscape, companies, citizens, and the urban infrastructure are part of this collective system.

As a system, complexity is one of the primary characteristics of transportation. As expressed by Cascetta, "[Transportation systems] are, indeed, internally complex systems, made up of many elements influencing each other both directly and indirectly, often nonlinearly, and with many feedback cycles." \cite[p.~v]{cascetta2009transportation}. The increasing complexity and dynamic nature of mobility make traditional planning fall short on the necessary flexibility and adaptability, and developing analytical optimization methods for transportation is no longer straightforward \cite{dorer2005adaptive}. This pushes us to think of new ways for planning that, as will be explained in following sections, could rely on self-organization capabilities of collaborative systems.

\subsection{Multi-functionality} \label{multi-functionality}

The complexity of urban mobility is partly due to the wide range of needs that it needs to cover. If we think of commuting trips, mobility systems solve the need of transporting people to and from home to work in the city. However, people also travel for leisure, emergencies, or just to run errands. Depending on their age and ability, people have different mobility needs.  Moreover, there is also the need to transport goods, which can vary in shapes, sizes, and urgency levels. Goods are sometimes delivered between people,  from a business to a person and vice versa, or between businesses. Lastly, urban mobility also covers services such as street cleaning, trash pickup, gardening, or infrastructure maintenance.

Going back to the cardiovascular system as a figurative metaphor for urban mobility and analysing the goals of both systems, the main goal of the cardiovascular system is to transport oxygen and metabolites through the body while collecting carbon dioxide and other wastes for removal \cite{quarteroni2000computational}. In urban mobility, there is also a need to transport people and goods and remove waste; however, in mobility these tasks are carried out separately. The cardiovascular system inspires us to think of multi-functional systems that can perform these tasks simultaneously. 

Today cars are stationary during 95\% of their lifetime \cite{hudda2013self} and, while it is estimated that sharing and autonomy will increase vehicle utilization rates, there will still be a representative percentage of the time, up to 50-80 \%, when the vehicles are not in use \cite{vosooghi2019shared, sanchez2021simulation}. During these periods, vehicles could be used for other purposes such as transporting goods or doing utility services, which could further increase the efficiency in the use of the vehicles and, consequently, reduce the number of vehicles in the streets. For instance, it already has been proposed to have lightweight electric vehicles that could transport people and goods at the same time \cite{lin2021affordable, yao2019idk}. In a similar way to how the cardiovascular system transports oxygen and metabolites while removes waste \cite{quarteroni2000computational}, we believe that having vehicles that are able to perform multiple functions will be essential for efficient transport. 

\subsection{Collaboration}\label{collaboration}

Collaboration, the third ingredient in this proposal, describes the way vehicles will interact in order to build a collective system. Taking nature as an example again, insects that live in colonies, such as ants, termites or honeybees show a collective behavior named natural swarms \cite{Gro2006SelfassemblyOM}. 

Based on collaboration natural swarms can divide tasks, cluster, build together, or transport cooperatively, among others \cite{bonabeau1999swarm}.  For instance ants can coordinate in colonies of hundreds of thousands of members with only local tactile and chemical communication \cite{deneubourg1989blind}. In some ant species, ants can find the food sources that are closest to the nest, as well as finding the minimum path between the nest and the food as a combination of the actions of many ants \cite{bonabeau1999swarm}. Ants are also able to transport large prey collaboratively; when working together, ants can move 5,000 times the weight and 10,000 times the volume of a single ant \cite{holldobler1978multiple} \cite{feinerman2018physics}.  In some species, ants chain their bodies forming bridges over large gaps so that other ants can walk over \cite{holldobler1978multiple}. In others, colonies are formed by ants of two different sizes; the smaller ones take care of the most quotidian tasks such as brooding or cleaning the nest, while the larger ones are specialized in milling seeds or defending the nest \cite{bonabeau1999swarm, wilson1984relation}.

Some of these behaviors have been adopted by the field of swarm robotics \cite{Beni2004FromSI}. If we translate these concepts from the behavior of ants into urban mobility, we could imagine urban mobility to work as a collective system; large fleets of vehicles might be able to cooperate based on only local communication, vehicles might be able to find the shortest paths though collective behavior without previous knowledge of the environment, vehicles might also be able to transport large objects collaboratively or specialize and divide tasks. All in all, based on these collaborative relationships, urban mobility would behave in a more similar way to a natural system that transports people and goods through the city in a self-organized and demand-responsive way. 

While it can be intuitive to imagine vehicles from the same company collaborating as part of a system, it may be more challenging to imagine vehicles from different companies collaborating. We would arguue that current mobility is mainly based on individualistic behaviors of drivers on the road who mostly compete for their own interests, resulting in inefficient and unsafe road conditions.  In nature, there are different types of symbiotic relationships that define the ways of "the living together of two or more organisms in close association" \cite[p.~49]{margulis1971symbiosis}. Amongst these, mutualism is specially interesting because it is based o relationships that benefit both parts involved and, therefore, there is a very clear incentive for them to collaborate.  Mutualistic relationships are not new in human society; they have been part of social interactions since ancient times, when humans started to trade and exchange resources \citep{chertow2007uncovering}. Throughout history, businesses have learnt how exchanging energy, materials or information can bring further benefits than those coming from the sum of each company acting alone \citep{chertow2000industrial}.  
We believe that a transition from competition relationships in the roads towards mutualistic relationships that create benefit-benefit situations would radically transform urban mobility for the better.

\section{Drawing ideas from swarm robotics and vehicle platooning}\label{from-swarms-and-platooning}

The concept of future mobility as \textit{multi-functional} vehicles behaving as a  \textit{bio-inspired collaborative system} combines, expands, and translates ideas from nature, swarm robotics, and vehicle platooning.  Natural swarm behaviors have served as an inspiration for researchers in robotics, leading to the field of swarm robotics \cite{Sahin2004SwarmRF}. While the  application is different from mobility, in the field of swarm robotics there are excellent examples of translating behaviors from natural to artificial systems. 

Within the field of mobility, research in vehicle platooning studies how vehicles can travel in coordination \cite{maiti2017conceptualization}. While our view aims to extend interactions between vehicles from coordination to various ways of collaboration, vehicle platooning can be seen as a first step towards this future.

This section explains how some of the concepts of swarm robotics might be transferred into the field of mobility and combined with vehicle platooning. The characteristics and benefits of these models are also outlined in this section, and, in doing so, we also provide details on the characteristics and potential benefits of the future mobility proposed in this paper. 

\subsection{From swarm robotics to mobility} \label{swarms}

The term 'swarm' in the context of robotics was used for the first time by Fukuda et al.\cite{Fukuda1988ApproachTT} and G. Beni \cite{Beni1988TheCO} in 1988. Fukuda studied how to design a robotic system that, having an intelligence analogous to the biological gene, would be able to dynamically reorganize its shape and structure for a given task by employing limited available resources \cite{Fukuda1988ApproachTT}. G. Beni, instead, defined what he named as Cellular Robotic System (CRS).

\begin{quotation}
The systems considered consist of a large (but finite) number of relatively simple robotic units capable of accomplishing, collectively, relatively complex tasks. ... The system formed by the autonomous robotic units called Cellular Robotic System or CRS is characterized by its reliability and its ability to self-organize and self-repair. \cite[p.~57]{Beni1988TheCO}.
\end{quotation}

In 1993, G. Beni and J. Wang \cite{10.1007/978-3-642-58069-7_38} argued the unpredictability in the behavior of systems that are capable of producing order can result in a non-trivial, different form of intelligence, which was named as 'swarm intelligence'. Swarm intelligence is a form of artificial intelligence inspired by insect colonies, emerging from the collective association of individual agents behaving in a way that goes beyond the aggregation of the individual capabilities of each agent \cite{innocente2019self}.

The field of swarm robotics implements swarm intelligence into multi-robot systems \cite{innocente2019self}. Trying to emulate the natural behavior of social insects, researchers in this field have pursued imitating behaviors like foraging, flocking, sorting, stigmergy, or cooperation \cite{Gro2006SelfassemblyOM}, which are based only on local information \cite{Gro2013TheSE}. Based on the same principle, swarm robotics aim to construct robust, flexible, and scalable systems by using simple robots and local interactions \cite{Gro2005AutonomousSI}.

Swarm robots are characterized by some properties such as simplicity, self-organization, decentralization,  task division and local interactions. These characteristics lead to scalability, robustness and flexibility, benefits that  are also critical to urban mobility systems. Both these properties and benefits have been defined by G. Beni \cite{Beni2004FromSI} and completed by others \cite{Sahin2004SwarmRF,Tan2013ResearchAI,Zakiev2018SwarmRR}. The following sections provide an overview of these characteristics indicating how they could impact mobility.

\subsubsection{Self-Organization}

In the same way that biological swarms self-organize into patterns, robotic swarms show intelligence by producing ordered structures in an unpredictable way \cite{Beni2004FromSI,Brambilla2012SwarmRA}. Self-organized systems in mobility can outperform traditional approaches based on fixed routes and schedules \cite{alfeo2019urban} and self-organization could solve some of the challenges inherent to vehicle networks \cite{Cherif2009ANF}.

\subsubsection{Decentralization}

Robots act by themselves when performing tasks with a distributed control topology and, consequently, any swarm member can make decisions independently \cite{Tan2013ResearchAI}. In connected mobility, decentralization can be critical to increase security and robustness, decreasing delays in communication, and solving the problem of a single node failure, as has been proved in research about network topologies \cite{Dorri2016BlockchainII}.  Decentralization could also improve scalability; having self-coordinating agents without a centralized communication allows having a greater number of agents \cite{innocente2019self}.

\subsubsection{Local Interactions}

Inter-robot communications and sensing are limited to be only local; a property directly inherited from natural swarms, which heavily rely on local interactions for cooperation \cite{Zakiev2018SwarmRR,Tan2013ResearchAI}. Local interactions lead to benefits such as flexibility, and robustness \cite{Brambilla2012SwarmRA}. In addition to these advantages, a system based on local interactions can be very scalable by dynamically adapting to different fleet sizes without any change in the software or hardware, which is very relevant for real-world applications such as urban mobility \cite{Tan2013ResearchAI}.

\subsubsection{Task division}

Robotic swarms act as a massive parallel computing system that, being able to create various solutions through task division between robots \cite{Cheraghi2021PastPA}, can carry out more complex tasks than the individual itself \cite{Tan2013ResearchAI}. This improves efficiency and performance in task completion as compared to single robotic systems \cite{Arkin1998AnBR}. 

\subsubsection{Simplicity}

Agents in a swarm are simple, which has been defined as having a limited capability relative to the global task \cite{innocente2019self}. Additionally, the robots are usually quasi-identical, which allows for standardization and cost efficiency \cite{Sahin2004SwarmRF}. Current autonomous vehicles are much less simple and more capable relative to the task than swarm robots; however, working in swarm systems could potentially decrease the level of complexity of autonomous vehicles.

\subsubsection{Scalability}

Thanks to local interactions and self-organization, robotic swarms could theoretically count on an unlimited number of members \cite{zakiev2018swarm}. In practice, robotic swarms could be composed of thousands, or even millions of units \cite{Beni2004FromSI}. While clusters of vehicles might be formed by a smaller number of vehicles, collaborative systems of two or a few agents could already show many of the characteristics seen in larger swarms, such as task division, decentralization, or local interactions, and so show some of its benefits \cite{Dorigo2013SwarmanoidAN}. 

\subsubsection{Robustness}

The number of agents in a swarm must be large enough to provide redundancy and robustness so that swarms can continue working even after the failure of some individuals \cite{Beni2004FromSI}. Robustness is also fundamental in transportation. Numerous studies have shown the relevance of reliability in mobility mode choice behavior \cite{carrion2012value,Cherif2009ANF}, showing that unreliability results in a negative effect in mode choice for commute trips, even for those who have flexible work schedules \cite{bhat2006impact}. Moreover, in the event of disasters, robustness in transportation is critical since it provides access to emergency supplies and services \cite{al2006new}, and it is also needed in order to restore other services such as water, electricity, or communications \cite{du1997degradable}.

\subsubsection{Flexibility}

Task division provides either optimization eliminating redundant efforts or extra security through redundancy \cite{Sahin2004SwarmRF}. For instance, a cluster of autonomous vehicles traveling together can share the navigation tasks required for autonomous driving, provide redundancy in the tasks that are more critical, or enable the vehicles to increase their performance thanks to the combined computational and physical capabilities.

In light of the above, there seems to be great potential in extending these concepts from swarm systems to mobility. The synergies of a future collaborative mobility would bring human behavior closer to natural systems, which have evolved through millions of years of evolution and are clear evidence of the power of collaboration \cite{Ahmed2012SwarmIC}.

\subsection{From vehicle platooning to collaboration}\label{platooning}

Autonomous vehicles can communicate with other vehicles, humans, and the infrastructure and have the intelligence to make decisions based on this communication \cite{harding2014vehicle}. These communication skills open the possibility for autonomous vehicles to cooperate, giving birth to concepts as vehicle platooning \cite{Kavathekar2011VehiclePA}.  Maiti et al. define vehicle platooning as a closely following mechanism that allows vehicles to travel in a coordinated way, without any mechanical linkage, while maintaining a safety distance \cite{maiti2017conceptualization}.

The concept of vehicle platooning has its origins in the 1960's \cite{Hanson1966ProjectM}. It started to gain attention when, in 1995, a research report related to the California PATH project showed promising results related to drag force reduction \cite{Zabat1995TheAP}. One of the most striking early  demonstrations of these concepts took place years later, in the 2011 Grand Cooperative Driving Challenge (GCDC) \cite{Ploeg2012IntroductionTT}, showing vehicles dynamically cooperating both in urban and highway scenarios. 

Vehicle platooning has three fundamental elements: 1) Vehicle-to-Vehicle (V2V) communication that allows vehicles to coordinate with minimal resources \cite{Shladover2006PATHA2,Bergenhem2012OVERVIEWOP}, 2) distributed control as a robust framework that allows each vehicle to decide how to act as a combination of the information received by the on-board sensors and V2V communication \cite{Bergenhem2012OVERVIEWOP,Alam2011FuelEfficientDC}, and 3) Vehicle-to-Infrastructure (V2I) communication which allows vehicles to interact with the infrastructure in a variety of scenarios such as finding a charging spot, or navigating to attach to a platoon \cite{Bergenhem2010CHALLENGESOP}. 

Platooning has multiple potential benefits, some of which are summarized in this section. Even if platooning research has mainly focused on heavy vehicles, extending this concept for lighter vehicles as we propose in this article could lead to similar benefits \cite{Shladover2006PATHA2}. Since platooning (i.e., coordinated travel) can be seen as a first step towards collaboration, the framework presented would inherit the benefits of platooning. These benefits are analyzed in the following sections. 

\subsubsection{Fuel Consumption}

The main research line in vehicle platooning is related to heavy vehicles due to direct benefits in fuel consumption associated with drag reductions \cite{Alam2011FuelEfficientDC,Tsugawa2011AnAT}. Drag reductions lead to environmental and financial gains, which is very relevant as fuel is the highest cost for a heavy vehicle fleet owner \cite{Bergenhem2012OVERVIEWOP}. 

\subsubsection{Traffic flow efficiency}

Systems based on velocity and space regulation such as the Intelligent Cruise Control System (ICC) have been proposed to increase traffic flow efficiency \cite{Darbha1998IntelligentCC}. For instance, in an early study Michael et al. \cite{Michael1998CapacityAO} quantified the increase in traffic flow efficiency, concluding that platoons formed by up to ten light-duty passenger vehicles could double or even triple highway capacity. 
However, must be noted that several studies indicate that, even if connected autonomous vehicles might increase traffic flow efficiency, this would, in turn, induce travel demand \cite{lee1999induced, hymel2010induced,narayanan2020factors}.

\subsubsection{Safety}

Among all traffic accidents, human error is involved in 50\% to 90\% of them \cite{Peters2002AutomotiveVS}. It is predicted that autonomous vehicle platooning will contribute to road safety with systems such as Cooperative Adaptive Cruise Control (CACC); a system in which vehicles constantly communicate and measure distances to their predecessors \cite{Axelsson2017SafetyIV, Arem2006TheIO}. Similarly to how driver assistance functions are becoming each time more advanced, CACC systems are also evolving towards more advanced predictive safety systems \cite{axelsson2016safety,van2019string}. %

The future mobility that is presented in this article extends the interactions between vehicles from the coordination proposed by platooning models to a broader set of interactions. Being an extended version of platooning, it would inherit the aforementioned benefits could potentially reduce fuel consumption, increase traffic flow efficiency and improve safety.

In addition to expanding the possible interactions between vehicles, this model also proposes a different approach to the way platoons would be formed. Most platooning research considers centralized planning approaches but it has been argued that platoons might be formed ad-hoc between drivers who do not necessarily know each other \cite{khan2005convoy}. Since in our proposed framework clusters of vehicles would be formed by vehicles that might even be from different fleet owners, it would also require decentralized planning strategies that would enable collaboration between vehicles to happen spontaneously and without previous planning, based on local interactions.

Lastly, while most platooning research considers that all the agents are identical or at least very similar \cite{Alam2010AnES}, proposal introduces the possibility of collaborating across scales and vehicle types forming heterogeneous clusters. We imagine that some vehicles could have specific duties such as charging or providing maintenance services to other vehicles. Taking this idea even a step further, vehicles could also different levels of autonomy as well; for instance, vehicles with outdoor driving capabilities could guide vehicles that can only navigate indoors from one building to another.

All in all, our proposal for future mobility builds on top of the research in vehicle platooning by bringing concepts from swarm robotics that expand interactions between vehicles from communication to a broader set of interactions that imitate the behavior found in natural swarms. Consequently, the potential benefits would also include those of swarm robotics and vehicle platooning. 

\section{Application to the main urban mobility needs} \label{towards-this-future}

The previous sections have provided a definition and a framework for a future mobility  from the perspective of an academic contribution. This section aims to convey our interpretation of this framework by including how we envision that these concepts could be applied to the main urban mobility needs: the movement of people, the movement of goods, and utility services. In doing so, we also outline some of the benefits of this possible future that add to those inherited from vehicle platooning and swarms.

\subsection{Movement of people} \label{people}

Traditional transport is not well suited to provide coverage of all regions, mobility needs, and population groups in an efficient and affordable way \cite{papanikolaou2017methodological}. For more than three decades, demand-responsive transport (DRT) has been regarded as a solution to this issue as alternative to conventional services such as buses and taxis for more than three decades \cite{mageean2003evaluation}. Its goal is to provide an efficient solution by combining the capacity of regular fixed transport services with the flexibility of on-demand services \cite{papanikolaou2017methodological}. However, current on-demand services such as Uber or Lyft that provide great flexibility, do not offer the capacity of traditional transport. 

In the framework of mobility as a bio-inspired collaborative system, the movement of people could occur as follows. Vehicles that shared some path could cluster for part of their trip and collaborate by, for instance, sharing battery, data, or computational power. Thanks to reduced aerodynamic drag forces, vehicles could save energy or travel faster, similarly to vehicle platoons \cite{Tsugawa2011AnAT}.  Then, when their paths diverged, vehicles would simply detach and continue on their own.  The way that this collective transportation would be generated is intrinsically demand-responsive since it would organically appear where vehicles are, as opposed to fixed-schedule and fixed-route services. As a demand-responsive solution for collective transportation, it could address the yet unmet goal of DRT of combining the flexibility of on-demand services with the capacity of traditional systems.

Another very relevant mobility-related concept that is more recent than DRT is Mobility as a Service (MaaS). MaaS aims to provide a 'seamless' multi-modal transportation with an access-based model by offering customized mobility packages through an integrated interface that provides services such as trip planning, reservations, and payment \cite{jittrapirom2017mobility, chowdhury2016users, nikitas2020artificial}. In this sense, a collaborative mobility framework could simultaneously offer the advantages of MaaS and mobility on-demand. Clustering individual mobility modes with other vehicles could create a seamless connection between a single-person mobility mode and a new collective transportation method. There would be no need for intermodality, and therefore, of transitioning between mobility modes, which would make the commuting experience would be seamless in a much higher degree than in MaaS.

\begin{figure}[h]
    \centering
    \includegraphics[width=4in]{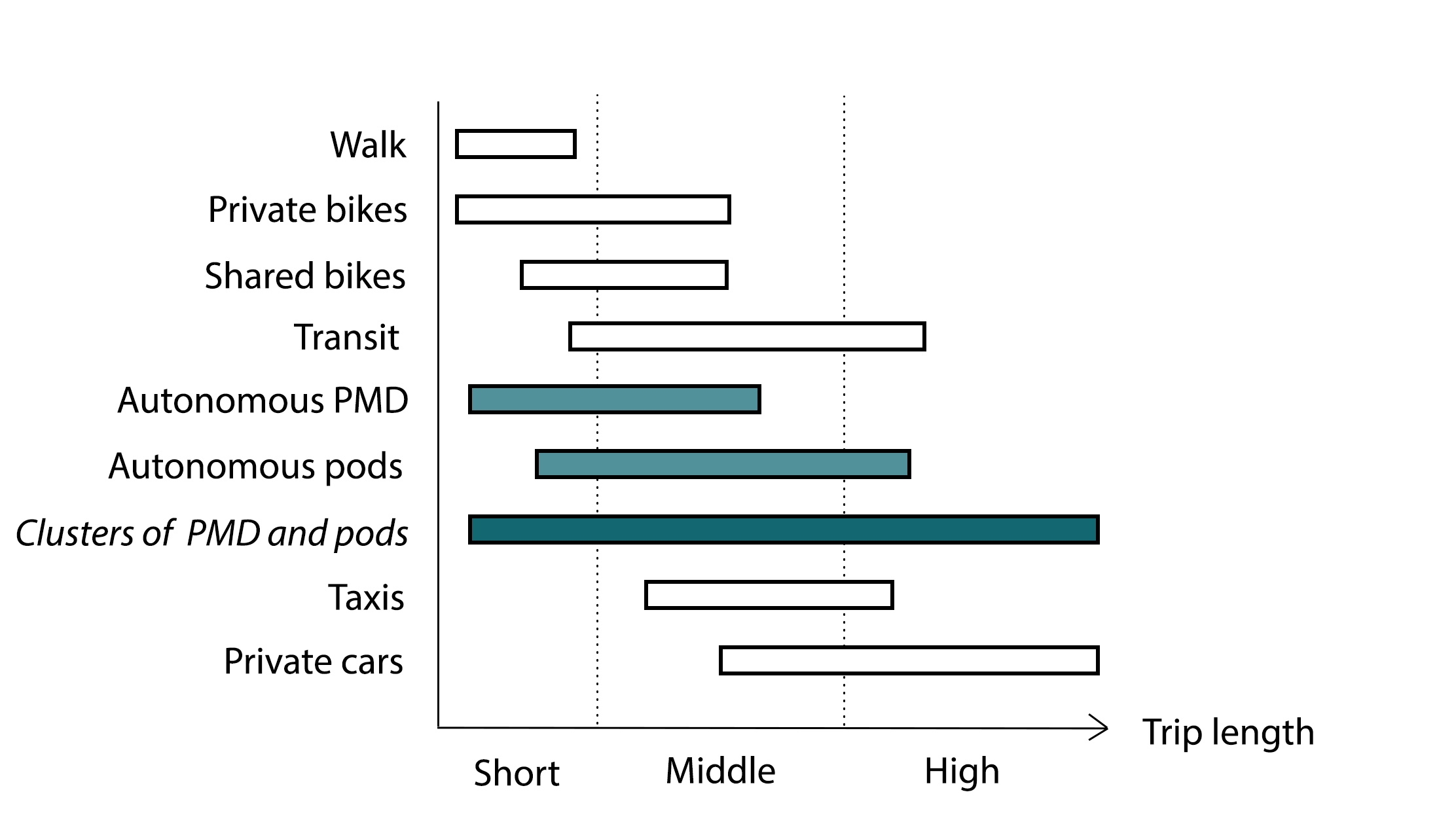}
    \caption{Trip length coverage of different mobility modes including autonomous personal mobility devices (PMDs) and autonomous pods, illustrating how clustering could extend the  trip lengths covered.  Adapted from \cite{curran2008}}
    \label{fig:fig3}
\end{figure}

Figure \ref{fig:fig3} represents the characteristics of the trips that would be covered by new mobility systems such as autonomous personal mobility devices (PMDs) which are single-person vehicles, and pods, which are typically four- or six-seaters. PMDs would provide a service in between shared bikes and private bikes and autonomous pods, instead, would provide a service in between transit, taxis, and private vehicles.  However, these systems are limited in capacity and the trip lengths that they cover.  Forming clusters could extend the trip lengths that vehicles serve, as well as their capacity. As a consequence, clusters of vehicles would provide a service that is more flexible than public transportation, without the need of heavy infrastructure, and with a capacity that is higher than the one offered by PMDs and pods.

All in all, a future in which vehicles would behave as a collaborative system would offer a mobility solution that combines the flexibility and convenience of door-to-door transportation with the efficiency of mass transit.  In addition, one could imagine many other possibilities that could be enabled by vehicle clustering in a collaborative system. For example, vehicles could have specialized tasks such as charging the vehicles, doing maintenance operations, or even providing services to the passengers such as in-transit food or beverages. Clustering could also provide a way for traveling in groups of family or friends using micro-mobility systems even when sharing just part of the trip and even if each person preferred to use different vehicles.

\subsection{Movement of goods}\label{goods}

Freight transport is an important source of economic growth and, at the same time, responsible for a substantial amount of $\mathrm{CO_2}$ emissions, traffic congestion, and road accidents \cite{Piecyk2010ForecastingTC}. In the future mobility scenario proposed in this paper, fixed-sized trucks could be substituted by clusters of smaller vehicles that would provide much greater flexibility since the capacity could be adapted to the need at each moment by just having more or fewer units, consequently optimizing resources.

For example, instead of having a truck deliver the cargo from the central warehouse to smaller warehouses, a van to transport it to a neighborhood scale, and a delivery person with a cart for the last metres, the entire process could be served by the same vehicles. These smaller vehicles would be clustered in a big number to move cargo in the metropolitan or even inter-metropolitan scale, then divide in smaller clusters to travel in the intrametropolitan scale, and finally separate into individual vehicles for the last-mile delivery in the neighborhood scale. 

The use of smaller and more lightweight vehicles could reduce fatalities since collisions with trucks are especially deadly \cite{george2017investigation}. Moreover, these smaller vehicles would provide the necessary agility for narrow streets with limited accessibility and maneuvering space, solving one of the most common issues of last-mile delivery \cite{alessandrini2015automated}.

Our proposed framework could also complement what is known as Collaborative Transport (CT). CT proposes the collaboration between companies as a way to optimize the global supply chain network to reduce costs \cite{gonzalez2012defining}. It is considered to be a very efficient approach to make transport more efficient and improve problems such as $\mathrm{CO_2}$ emissions or congestion and has attracted great research interest in recent years \cite{piecyk2010forecasting, Pan2019HorizontalCT}. There are two main types of CT: vertical collaborative transport (VCT), where partners serve distinct parts of the network and horizontal collaborative transport (HCT), in which partners cover the same or at least overlapping segments of the supply chain \cite{Cleophas2019CollaborativeUT}. Our proposed framework can be an interesting complement for CT.  Thanks to the flexibility in capacity, vehicle clustering could extend the possibilities of creating HCT models that simultaneously cover more segments of the supply chain.  As our proposed collaborative mobility would have decentralized planning like swarm systems and it could work even with limited exchange of information \cite{elbert2020analysis}, it could provide the privacy that CT needs to ensure the protection of companies' sensitive information and decision-making competencies \cite{Wang2014CollaborativeTP}. Finally, if clustering would allow vehicles to transfer cargo to other vehicles while moving, it would open yet another set of opportunities in the design of collaborative transportation systems.

Summarizing, the resulting system would be composed of small and agile vehicles which, together, could have enough loading capacity to cover the city's needs substituting heavy vehicles that, in addition to their lack of flexibility and maneuverability, pose a threat to pedestrians.

\subsection{Utility services}\label{utility-services}

Due to the increasing speed at which cities grow, the management of utility services is becoming increasingly important \cite{zhang2014visual}. In fact, it is estimated that the global investment needed in infrastructure will be 3.7 trillion USD per year through 2035, and 1 trillion USD per year more to meet the United Nations Sustainable Development Goals \cite{woetzel2017bridging}. This indicates very clearly the need for rethinking urban infrastructure, and, therefore, how utility services are carried out. 

Currently, companies are developing autonomous robots for utility services such as surveillance, snow removal, and gardening (e.g., SMP Security robot, Snowbot, Yardroid). We believe that, in the future, swarm systems of these smaller robots will complete utility services collaboratively and replace traditional approaches such as large trucks or vans. 

Taking waste collection as an example, in previous research our research group has modeled a fleet of lightweight autonomous Persuasive Electric Vehicles (PEVs) behaving like a swarm system that picked up the trash in the area of Kendall, Cambridge (USA) \cite{alfeo2019urban}. The PEV is an autonomous tricycle developed by the by our lab that aims to provide a lightweight solution for the delivery of people and goods. In that study, PEVs was modeled to behave as a bio-inspired swarm, using methods such as multi-place foraging and stigmergy to collaboratively carry waste from trash bins on the street to a central deposit. The study showed that this self-organized system outperformed the current approach that uses traditional garbage trucks, reducing both the average amount of trash on the trash bins and the average number of full trash bins \cite{alfeo2019urban}. 

We believe that the same approach could be applied for other utility services, replacing traditional systems with fleets of smaller vehicles that would provide a more efficient service by working collaboratively.

\section{Conclusions} \label{conclusions}

In the current trend towards electrification, autonomy, and sharing, little attention is being paid to what will come after these trends are settled, specially from the perspective of interactions between vehicles and how these interactions might affect the system-level behavior. Mobility is an increasingly complex system, and, in nature, there are various examples of complex systems with similarities to mobility that have evolved to be collaborative. Research in swarm robotics has demonstrated the potential benefits of translating these natural behaviors into artificial systems. Moreover, research in vehicle platooning has demonstrated that coordination between vehicles can lead to more efficient traffic, reduced energy consumption, and increased safety. 

This paper proposes a future that combines, expands, and translates ideas from nature, swarm robotics, and vehicle platooning, proposing a future mobility based on three key ingredients: system behavior, multi-functionality and collaboration. In addition to inheriting the benefits of vehicle platooning and swarms, this proposal can radically transform the way cities address the three main mobility needs: the movement of people, movement of goods, and utility services.

The framework proposed in this paper points towards a future in which mobility would work more similarly to how natural systems behave. Since natural processes have been tailored by millions of years of evolution, natural systems are more efficient, sustainable, and resilient, which are characteristics that we want our future transportation systems to have.

\section*{Disclosure statement}

No potential competing interest was reported by the authors.

 \bibliographystyle{elsarticle-num} 
 \bibliography{cas-refs}

\begin{thebibliography}{100}
\expandafter\ifx\csname url\endcsname\relax
  \def\url#1{\texttt{#1}}\fi
\expandafter\ifx\csname urlprefix\endcsname\relax\def\urlprefix{URL }\fi
\expandafter\ifx\csname href\endcsname\relax
  \def\href#1#2{#2} \def\path#1{#1}\fi

\bibitem{EPA2021}
{United States Environmental Protection Agency (EPA)}, Inventory of u.s.
  greenhouse gas emissions and sinks 1990–2019 (2021).

\bibitem{UNreport2018}
{United Nations Department of Economic and Social Affairs Population Division},
  World urbanization prospects: The 2018 revision (2019).

\bibitem{McKinsey2019}
T.~Möller, A.~Padhi, D.~Pinner, A.~Tschiesner, The future of mobility is at
  our doorstep (2019).

\bibitem{alarfaj2020decarbonizing}
A.~F. Alarfaj, W.~M. Griffin, C.~Samaras, Decarbonizing us passenger vehicle
  transport under electrification and automation uncertainty has a travel
  budget, Environmental Research Letters 15~(9) (2020) 0940c2.

\bibitem{zhang2020role}
R.~Zhang, S.~Fujimori, The role of transport electrification in global climate
  change mitigation scenarios, Environmental Research Letters 15~(3) (2020)
  034019.

\bibitem{martinez2015urban}
L.~Martinez, P.~Crist, Urban mobility system upgrade--how shared self-driving
  cars could change city traffic, in: International Transport Forum, Paris,
  2015.

\bibitem{martin2010impact}
E.~Martin, S.~A. Shaheen, J.~Lidicker, Impact of carsharing on household
  vehicle holdings: Results from north american shared-use vehicle survey,
  Transportation research record 2143~(1) (2010) 150--158.

\bibitem{millard2005car}
A.~Millard-Ball, Car-sharing: Where and how it succeeds, Vol.~60,
  Transportation Research Board, 2005.

\bibitem{shaheen1999short}
S.~A. Shaheen, D.~Sperling, C.~Wagner, A short history of carsharing in the
  90's (1999).

\bibitem{he2017service}
L.~He, H.-Y. Mak, Y.~Rong, Z.-J.~M. Shen, Service region design for urban
  electric vehicle sharing systems, Manufacturing \& Service Operations
  Management 19~(2) (2017) 309--327.

\bibitem{jones2019contributions}
E.~C. Jones, B.~D. Leibowicz, Contributions of shared autonomous vehicles to
  climate change mitigation, Transportation Research Part D: Transport and
  Environment 72 (2019) 279--298.

\bibitem{greenblatt2015autonomous}
J.~B. Greenblatt, S.~Saxena, Autonomous taxis could greatly reduce
  greenhouse-gas emissions of us light-duty vehicles, Nature Climate Change
  5~(9) (2015) 860--863.

\bibitem{narayanan2020shared}
S.~Narayanan, E.~Chaniotakis, C.~Antoniou, Shared autonomous vehicle services:
  A comprehensive review, Transportation Research Part C: Emerging Technologies
  111 (2020) 255--293.

\bibitem{cervero2017mobility}
R.~Cervero, Mobility niches: jitneys to robo-taxis, Journal of the american
  planning association 83~(4) (2017) 404--412.

\bibitem{sanchez2020autonomous}
N.~C. Sanchez, L.~A. Pastor, K.~Larson, Autonomous bicycles: A new approach to
  bicycle-sharing systems, in: 2020 ieee 23rd international conference on
  intelligent transportation systems (itsc), IEEE, 2020, pp. 1--6.

\bibitem{milovanoff2020electrification}
A.~Milovanoff, I.~D. Posen, H.~L. MacLean, Electrification of light-duty
  vehicle fleet alone will not meet mitigation targets, Nature Climate Change
  10~(12) (2020) 1102--1107.

\bibitem{EIACarbonIntensity}
{U.S. Energy Information Administration}, How much carbon dioxide is produced
  per kilowatthour of u.s. electricity generation?,
  \url{https://www.eia.gov/tools/faqs/faq.php?id=74\&t=11} (2021).

\bibitem{EEA}
{European Environment Agency (EEA)}, Greenhouse gas emission intensity of
  electricity generation by country,
  \url{https://www.eea.europa.eu/data-and-maps/daviz/co2-emission-intensity-9/}
  (2021).

\bibitem{ewing2010travel}
R.~Ewing, R.~Cervero, Travel and the built environment: A meta-analysis,
  Journal of the American planning association 76~(3) (2010) 265--294.

\bibitem{alonso2018cityscope}
L.~Alonso, Y.~R. Zhang, A.~Grignard, A.~Noyman, Y.~Sakai, M.~ElKatsha,
  R.~Doorley, K.~Larson, Cityscope: a data-driven interactive simulation tool
  for urban design. use case volpe, in: International conference on complex
  systems, Springer, 2018, pp. 253--261.

\bibitem{grignard2018cityscope}
A.~Grignard, N.~Maci{\`a}, L.~Alonso~Pastor, A.~Noyman, Y.~Zhang, K.~Larson,
  Cityscope andorra: a multi-level interactive and tangible agent-based
  visualization, in: Proceedings of the 17th International Conference on
  Autonomous Agents and MultiAgent Systems, 2018, pp. 1939--1940.

\bibitem{lin2021affordable}
M.~C.-L. Lin, Affordable autonomous lightweight personal mobility, Ph.D.
  thesis, Massachusetts Institute of Technology (2021).

\bibitem{yao2019idk}
J.~W.-H. Yao, Idk: an interaction development kit to design interactions for
  lightweight autonomous vehicles, Ph.D. thesis, Massachusetts Institute of
  Technology (2019).

\bibitem{harding2014vehicle}
J.~Harding, G.~Powell, R.~Yoon, J.~Fikentscher, C.~Doyle, D.~Sade, M.~Lukuc,
  J.~Simons, J.~Wang, et~al., Vehicle-to-vehicle communications: readiness of
  v2v technology for application., Tech. rep., United States. National Highway
  Traffic Safety Administration (2014).

\bibitem{maiti2017conceptualization}
S.~Maiti, S.~Winter, L.~Kulik, A conceptualization of vehicle platoons and
  platoon operations, Transportation Research Part C: Emerging Technologies 80
  (2017) 1--19.

\bibitem{bonabeau1999swarm}
E.~Bonabeau, D.~d. R. D.~F. Marco, M.~Dorigo, G.~Th{\'e}raulaz, G.~Theraulaz,
  et~al., Swarm intelligence: from natural to artificial systems, no.~1, Oxford
  university press, 1999.

\bibitem{Sahin2004SwarmRF}
E.~Sahin, Swarm robotics: From sources of inspiration to domains of
  application, in: Swarm Robotics, 2004.

\bibitem{gonzalez2021dynamic}
J.~M. Gonz{\'a}lez, N.~C. S{\'a}nchez, P.~G.~M. Llop, L.~A. Pastor, K.~Larson,
  Dynamic v2v power-sharing for collaborative fleets of autonomous vehicles,
  in: 2021 IEEE Electrical Power and Energy Conference (EPEC), IEEE, 2021, pp.
  408--413.

\bibitem{cruz2015wrongfully}
P.~Cruz, Wrongfully right: applications of semantic figurative metaphors in
  information visualization, IEEE VIS Arts Program (VISAP) (2015) 14--21.

\bibitem{stefanovska1999physics}
A.~Stefanovska, Physics of the human cardiovascular system, Contemporary
  Physics 40~(1) (1999) 31--55.

\bibitem{cascetta2009transportation}
E.~Cascetta, Transportation systems analysis: models and applications, Vol.~29,
  Springer Science \& Business Media, 2009.

\bibitem{dorer2005adaptive}
K.~Dorer, M.~Calisti, An adaptive solution to dynamic transport optimization,
  in: Proceedings of the fourth international joint conference on Autonomous
  agents and multiagent systems, 2005, pp. 45--51.

\bibitem{quarteroni2000computational}
A.~Quarteroni, M.~Tuveri, A.~Veneziani, Computational vascular fluid dynamics:
  problems, models and methods, Computing and Visualization in Science 2~(4)
  (2000) 163--197.

\bibitem{hudda2013self}
R.~Hudda, C.~Kelly, G.~Long, J.~Luo, A.~Pandit, D.~Phillips, I.~Sidhu, Self
  driving cars, College of Engineering University of California, Berkeley,
  Berkeley: College of Engineering University of California (2013).

\bibitem{vosooghi2019shared}
R.~Vosooghi, J.~Puchinger, M.~Jankovic, A.~Vouillon, Shared autonomous vehicle
  simulation and service design, Transportation Research Part C: Emerging
  Technologies 107 (2019) 15--33.

\bibitem{sanchez2021simulation}
N.~C. S{\'a}nchez, I.~Martinez, L.~A. Pastor, K.~Larson, Simulation study on
  the fleet performance of shared autonomous bicycles, arXiv preprint
  arXiv:2106.09694 (2021).

\bibitem{Gro2006SelfassemblyOM}
R.~Gro{\ss}, M.~Dorigo, M.~Yamakita, Self-assembly of mobile robots: From
  swarm-bot to super-mechano colony, in: IAS, 2006.

\bibitem{deneubourg1989blind}
J.-L. Deneubourg, S.~Goss, N.~Franks, J.~Pasteels, The blind leading the blind:
  modeling chemically mediated army ant raid patterns, Journal of insect
  behavior 2~(5) (1989) 719--725.

\bibitem{holldobler1978multiple}
B.~H{\"o}lldobler, E.~O. Wilson, The multiple recruitment systems of the
  african weaver ant oecophylla longinoda (latreille)(hymenoptera: Formicidae),
  Behavioral Ecology and Sociobiology 3~(1) (1978) 19--60.

\bibitem{feinerman2018physics}
O.~Feinerman, I.~Pinkoviezky, A.~Gelblum, E.~Fonio, N.~S. Gov, The physics of
  cooperative transport in groups of ants, Nature Physics 14~(7) (2018)
  683--693.

\bibitem{wilson1984relation}
E.~O. Wilson, The relation between caste ratios and division of labor in the
  ant genus pheidole (hymenoptera: Formicidae), Behavioral Ecology and
  Sociobiology 16~(1) (1984) 89--98.

\bibitem{Beni2004FromSI}
G.~Beni, From swarm intelligence to swarm robotics, in: Swarm Robotics, 2004.

\bibitem{margulis1971symbiosis}
L.~Margulis, Symbiosis and evolution, Scientific American 225~(2) (1971)
  48--61.

\bibitem{chertow2007uncovering}
M.~R. Chertow, “uncovering” industrial symbiosis, Journal of industrial
  Ecology 11~(1) (2007) 11--30.

\bibitem{chertow2000industrial}
M.~R. Chertow, Industrial symbiosis: literature and taxonomy, Annual review of
  energy and the environment 25~(1) (2000) 313--337.

\bibitem{Fukuda1988ApproachTT}
T.~Fukuda, S.~Nakagawa, Approach to the dynamically reconfigurable robotic
  system, Journal of Intelligent and Robotic Systems 1 (1988) 55--72.

\bibitem{Beni1988TheCO}
G.~Beni, The concept of cellular robotic system, Proceedings IEEE International
  Symposium on Intelligent Control 1988 (1988) 57--62.

\bibitem{10.1007/978-3-642-58069-7_38}
G.~Beni, J.~Wang, Swarm intelligence in cellular robotic systems, in: P.~Dario,
  G.~Sandini, P.~Aebischer (Eds.), Robots and Biological Systems: Towards a New
  Bionics?, Springer Berlin Heidelberg, Berlin, Heidelberg, 1993, pp. 703--712.

\bibitem{innocente2019self}
M.~S. Innocente, P.~Grasso, Self-organising swarms of firefighting drones:
  Harnessing the power of collective intelligence in decentralised multi-robot
  systems, Journal of Computational Science 34 (2019) 80--101.

\bibitem{Gro2013TheSE}
R.~Gro{\ss}, R.~O'Grady, A.~Christensen, M.~Dorigo, S.~Kernbach, The swarm-bot
  experience: Strength and mobility through physical cooperation, 2013.

\bibitem{Gro2005AutonomousSI}
R.~Gro{\ss}, M.~Bonani, F.~Mondada, M.~Dorigo, Autonomous self-assembly in a
  swarm-bot, in: AMiRE, 2005.

\bibitem{Tan2013ResearchAI}
Y.~Tan, Z.~Zheng, Research advance in swarm robotics, Defence Technology 9
  (2013) 18--39.

\bibitem{Zakiev2018SwarmRR}
A.~Zakiev, T.~Tsoy, E.~Magid, Swarm robotics: Remarks on terminology and
  classification, in: ICR, 2018.

\bibitem{Brambilla2012SwarmRA}
M.~Brambilla, E.~Ferrante, M.~Birattari, M.~Dorigo, Swarm robotics: a review
  from the swarm engineering perspective, Swarm Intelligence 7 (2012) 1--41.

\bibitem{alfeo2019urban}
A.~L. Alfeo, E.~C. Ferrer, Y.~L. Carrillo, A.~Grignard, L.~A. Pastor, D.~T.
  Sleeper, M.~G. Cimino, B.~Lepri, G.~Vaglini, K.~Larson, et~al., Urban swarms:
  A new approach for autonomous waste management, in: 2019 International
  Conference on Robotics and Automation (ICRA), IEEE, 2019, pp. 4233--4240.

\bibitem{Cherif2009ANF}
M.~Cherif, S.~Senouci, B.~Ducourthial, A new framework of self-organization of
  vehicular networks, 2009 Global Information Infrastructure Symposium (2009)
  1--6.

\bibitem{Dorri2016BlockchainII}
A.~Dorri, S.~Kanhere, R.~Jurdak, Blockchain in internet of things: Challenges
  and solutions, ArXiv abs/1608.05187 (2016).

\bibitem{Cheraghi2021PastPA}
A.~Cheraghi, S.~Shahzad, K.~Graffi, Past, present, and future of swarm
  robotics, ArXiv abs/2101.00671 (2021).

\bibitem{Arkin1998AnBR}
R.~Arkin, An behavior-based robotics, 1998.

\bibitem{zakiev2018swarm}
A.~Zakiev, T.~Tsoy, E.~Magid, Swarm robotics: remarks on terminology and
  classification, in: International Conference on Interactive Collaborative
  Robotics, Springer, 2018, pp. 291--300.

\bibitem{Dorigo2013SwarmanoidAN}
M.~Dorigo, D.~Floreano, L.~Gambardella, F.~Mondada, S.~Nolfi, T.~Baaboura,
  M.~Birattari, M.~Bonani, M.~Brambilla, A.~Brutschy, D.~Burnier, A.~Campo,
  A.~L. Christensen, A.~Decugniere, G.~D. Caro, F.~Ducatelle, E.~Ferrante,
  A.~F{\"o}rster, J.~M. Gonzales, J.~Guzzi, V.~Longchamp, S.~Magnenat,
  N.~Mathews, M.~M.~D. Oca, R.~O'Grady, C.~Pinciroli, G.~Pini, P.~R{\'e}tornaz,
  J.~Roberts, V.~Sperati, T.~Stirling, A.~Stranieri, T.~St{\"u}tzle,
  V.~Trianni, E.~Tuci, A.~E. Turgut, F.~Vaussard, Swarmanoid: A novel concept
  for the study of heterogeneous robotic swarms, IEEE Robotics \& Automation
  Magazine 20 (2013) 60--71.

\bibitem{carrion2012value}
C.~Carrion, D.~Levinson, Value of travel time reliability: A review of current
  evidence, Transportation research part A: policy and practice 46~(4) (2012)
  720--741.

\bibitem{bhat2006impact}
C.~R. Bhat, R.~Sardesai, The impact of stop-making and travel time reliability
  on commute mode choice, Transportation Research Part B: Methodological 40~(9)
  (2006) 709--730.

\bibitem{al2006new}
H.~Al-Deek, E.~B. Emam, New methodology for estimating reliability in
  transportation networks with degraded link capacities, Journal of intelligent
  transportation systems 10~(3) (2006) 117--129.

\bibitem{du1997degradable}
Z.-P. Du, A.~Nicholson, Degradable transportation systems: sensitivity and
  reliability analysis, Transportation Research Part B: Methodological 31~(3)
  (1997) 225--237.

\bibitem{Ahmed2012SwarmIC}
H.~Ahmed, J.~Glasgow, Swarm intelligence: Concepts, models and applications,
  2012.

\bibitem{Kavathekar2011VehiclePA}
P.~Kavathekar, Y.~Chen, Vehicle platooning: A brief survey and categorization,
  2011.

\bibitem{Hanson1966ProjectM}
M.~Hanson, Project metran : an integrated, evolutionary transportation system
  for urban areas, 1966.

\bibitem{Zabat1995TheAP}
M.~Zabat, N.~Stabile, S.~Farascaroli, F.~Browand, The aerodynamic performance
  of platoons: A final report, PATH research report (1995).

\bibitem{Ploeg2012IntroductionTT}
J.~Ploeg, S.~Shladover, H.~Nijmeijer, N.~V.~D. Wouw, Introduction to the
  special issue on the 2011 grand cooperative driving challenge, IEEE
  Transactions on Intelligent Transportation Systems 13 (2012) 989--993.

\bibitem{Shladover2006PATHA2}
S.~Shladover, Path at 20—history and major milestones, IEEE Transactions on
  Intelligent Transportation Systems 8 (2006) 584--592.

\bibitem{Bergenhem2012OVERVIEWOP}
C.~Bergenhem, S.~Shladover, E.~Coelingh, C.~Englund, S.~Tsugawa, Overview of
  platooning systems, 2012.

\bibitem{Alam2011FuelEfficientDC}
A.~Alam, Fuel-efficient distributed control for heavy duty vehicle platooning,
  2011.

\bibitem{Bergenhem2010CHALLENGESOP}
C.~Bergenhem, Challenges of platooning on public motorways, 2010.

\bibitem{Tsugawa2011AnAT}
S.~Tsugawa, S.~Kato, K.~Aoki, An automated truck platoon for energy saving,
  2011 IEEE/RSJ International Conference on Intelligent Robots and Systems
  (2011) 4109--4114.

\bibitem{Darbha1998IntelligentCC}
S.~Darbha, K.~Rajagopal, Intelligent cruise control systems and traffic flow
  stability, PATH research report (1998).

\bibitem{Michael1998CapacityAO}
J.~Michael, D.~Godbole, J.~Lygeros, R.~Sengupta, Capacity analysis of traffic
  flow over a single-lane automated highway system, J. Intell. Transp. Syst. 4
  (1998) 49--80.

\bibitem{lee1999induced}
D.~B. Lee~Jr, L.~A. Klein, G.~Camus, Induced traffic and induced demand,
  Transportation Research Record 1659~(1) (1999) 68--75.

\bibitem{hymel2010induced}
K.~M. Hymel, K.~A. Small, K.~Van~Dender, Induced demand and rebound effects in
  road transport, Transportation Research Part B: Methodological 44~(10) (2010)
  1220--1241.

\bibitem{narayanan2020factors}
S.~Narayanan, E.~Chaniotakis, C.~Antoniou, Factors affecting traffic flow
  efficiency implications of connected and autonomous vehicles: A review and
  policy recommendations, Advances in Transport Policy and Planning 5 (2020)
  1--50.

\bibitem{Peters2002AutomotiveVS}
G.~Peters, B.~Peters, Automotive vehicle safety, 2002.

\bibitem{Axelsson2017SafetyIV}
J.~Axelsson, Safety in vehicle platooning: A systematic literature review, IEEE
  Transactions on Intelligent Transportation Systems 18 (2017) 1033--1045.

\bibitem{Arem2006TheIO}
B.~Arem, C.~V. Driel, R.~Visser, The impact of cooperative adaptive cruise
  control on traffic-flow characteristics, IEEE Transactions on Intelligent
  Transportation Systems 7 (2006) 429--436.

\bibitem{axelsson2016safety}
J.~Axelsson, Safety in vehicle platooning: A systematic literature review, IEEE
  Transactions on Intelligent Transportation Systems 18~(5) (2016) 1033--1045.

\bibitem{van2019string}
E.~van Nunen, J.~Reinders, E.~Semsar-Kazerooni, N.~Van De~Wouw, String stable
  model predictive cooperative adaptive cruise control for heterogeneous
  platoons, IEEE Transactions on Intelligent Vehicles 4~(2) (2019) 186--196.

\bibitem{khan2005convoy}
M.~A. Khan, L.~Boloni, Convoy driving through ad-hoc coalition formation, in:
  11th IEEE real time and embedded technology and applications symposium, IEEE,
  2005, pp. 98--105.

\bibitem{Alam2010AnES}
A.~Alam, A.~Gattami, K.~Johansson, An experimental study on the fuel reduction
  potential of heavy duty vehicle platooning, 13th International IEEE
  Conference on Intelligent Transportation Systems (2010) 306--311.

\bibitem{papanikolaou2017methodological}
A.~Papanikolaou, S.~Basbas, G.~Mintsis, C.~Taxiltaris, A methodological
  framework for assessing the success of demand responsive transport (drt)
  services, Transportation Research Procedia 24 (2017) 393--400.

\bibitem{mageean2003evaluation}
J.~Mageean, J.~D. Nelson, The evaluation of demand responsive transport
  services in europe, Journal of Transport Geography 11~(4) (2003) 255--270.

\bibitem{jittrapirom2017mobility}
P.~Jittrapirom, V.~Caiati, A.-M. Feneri, S.~Ebrahimigharehbaghi, M.~J.
  Alonso~Gonz{\'a}lez, J.~Narayan, Mobility as a service: A critical review of
  definitions, assessments of schemes, and key challenges (2017).

\bibitem{chowdhury2016users}
S.~Chowdhury, A.~A. Ceder, Users’ willingness to ride an integrated
  public-transport service: A literature review, Transport Policy 48 (2016)
  183--195.

\bibitem{nikitas2020artificial}
A.~Nikitas, K.~Michalakopoulou, E.~T. Njoya, D.~Karampatzakis, Artificial
  intelligence, transport and the smart city: Definitions and dimensions of a
  new mobility era, Sustainability 12~(7) (2020) 2789.

\bibitem{curran2008}
A.~Curran, Translink public bike system feasibility study, Quay Communications
  Inc., Vancouver (2008).

\bibitem{Piecyk2010ForecastingTC}
M.~Piecyk, A.~McKinnon, Forecasting the carbon footprint of road freight
  transport in 2020, International Journal of Production Economics 128 (2010)
  31--42.

\bibitem{george2017investigation}
Y.~George, T.~Athanasios, P.~George, Investigation of road accident severity
  per vehicle type, Transportation research procedia 25 (2017) 2076--2083.

\bibitem{alessandrini2015automated}
A.~Alessandrini, A.~Campagna, P.~Delle~Site, F.~Filippi, L.~Persia, Automated
  vehicles and the rethinking of mobility and cities, Transportation Research
  Procedia 5 (2015) 145--160.

\bibitem{gonzalez2012defining}
J.~Gonzalez-Feliu, J.-M. Salanova, Defining and evaluating collaborative urban
  freight transportation systems, Procedia-Social and Behavioral Sciences 39
  (2012) 172--183.

\bibitem{piecyk2010forecasting}
M.~I. Piecyk, A.~C. McKinnon, Forecasting the carbon footprint of road freight
  transport in 2020, International Journal of Production Economics 128~(1)
  (2010) 31--42.

\bibitem{Pan2019HorizontalCT}
S.~Pan, D.~Trentesaux, E.~Ballot, G.~Huang, Horizontal collaborative transport:
  survey of solutions and practical implementation issues, International
  Journal of Production Research 57 (2019) 5340 -- 5361.

\bibitem{Cleophas2019CollaborativeUT}
C.~Cleophas, C.~Cottrill, J.~Ehmke, K.~Tierney, Collaborative urban
  transportation: Recent advances in theory and practice, Eur. J. Oper. Res.
  273 (2019) 801--816.

\bibitem{elbert2020analysis}
R.~Elbert, J.-K. Knigge, A.~Friedrich, Analysis of decentral platoon planning
  possibilities in road freight transport using an agent-based simulation
  model, Journal of Simulation 14~(1) (2020) 64--75.

\bibitem{Wang2014CollaborativeTP}
X.~Wang, H.~Kopfer, Collaborative transportation planning of
  less-than-truckload freight, OR Spectrum 36 (2014) 357--380.

\bibitem{zhang2014visual}
J.~Zhang, E.~Yanli, J.~Ma, Y.~Zhao, B.~Xu, L.~Sun, J.~Chen, X.~Yuan, Visual
  analysis of public utility service problems in a metropolis, IEEE
  transactions on visualization and computer graphics 20~(12) (2014)
  1843--1852.

\bibitem{woetzel2017bridging}
J.~Woetzel, N.~Garemo, J.~Mischke, P.~Kamra, R.~Palter, Bridging infrastructure
  gaps: Has the world made progress, McKinsey \& Company 5 (2017).

\end{thebibliography}





\end{document}